# Real-Space Dynamic Electron Correlation in Beryllium

Rudra B. Bista,[1] Yuya Shinohara,[2,*] Wojciech Dmowski,[3] Chae Woo Ryu,[3,†] Jung Ho Kim,[4] Mary Upton,[4] Hlynur Gretarsson,[5] Martin Sundermann,[5] and Takeshi Egami[1,2,3,‡]

[1]*Department of Physics and Astronomy, The University of Tennessee, Knoxville, Knoxville, Tennessee 37996, USA*
[2]*Materials Science and Technology Division, Oak Ridge National Laboratory, Oak Ridge, Tennessee 37830, USA*
[3]*Department of Materials Science and Engineering, The University of Tennessee, Knoxville, Knoxville, Tennessee 37996, USA*
[4]*X-ray Science Division, Argonne National Laboratory, Lemont, Illinois 60439, USA*
[5]*Deutsches Elektronen-Synchrotron DESY, 22607 Hamburg, Germany*

Electron correlation in solids has a major impact on material properties. However, it has been studied mainly by theory, with very limited direct experimental investigations. Here, we demonstrate that dynamic electron correlation function can be experimentally measured using inelastic x-ray scattering on polycrystalline beryllium. The data are expressed as the energy-resolved dynamic pair-distribution function. Our results confirm the size of the exchange-correlation hole as ∼2 Å, consistent with theoretical expectations. However, at the plasmon energy of ∼21 eV, the exchange-correlation hole is extended up to 4–5 Å, suggesting a unique influence of the dynamic plasmon state.

The nearly free electron theories in condensed matter physics are based on the idea that most fundamental physical phenomena in solids could be understood within a single-particle framework [1–3], even though electrons are dynamically correlated due to Coulomb interaction and constrained by the Pauli exclusion principle. This picture was supported by the Landau quasiparticle theory [4] and the density functional theory (DFT) [5]. Although the DFT has limitations in describing strongly correlated electron systems, which require many-body approaches such as the Hubbard model [3], it remains remarkably effective in explaining various phenomena in modern physics [6] and chemistry [7]. Electron correlation plays a central role in determining chemical bonding and material properties, and is likely a critical factor underlying emergent phenomena, such as superconductivity and charge ordering [8]. It also enables advanced applications including high-order harmonic generation [9], nanoscale single-electron sources [10], and single charge-based logic gates [11], making it a cornerstone of modern science and technology.

*Contact author: shinoharay@ornl.gov
†Present address: Department of Materials Science and Engineering, Hongik University, Seoul 04066, South Korea.
‡Contact author: egami@utk.edu

The DFT concept is based on the idea of exchange-correlation hole [12–14]. Despite its significance, direct experimental observations of exchange-correlation hole remain scarce. Attempts have been made to observe exchange-correlation hole through the electronic pair-distribution function (PDF), which is obtained by Fourier-transforming the total electronic structure function, $S(Q)$, where $Q$ is the momentum transfer in scattering, measured by X-ray diffraction [15–19]. However, a diffraction measurement is designed for elastic scattering, whereas information on electron correlation is contained only in inelastic scattering, because the elastic scattering intensity reflects only the spatial variations in static electron density. This mismatch results in inaccuracies. In diffraction experiment a detector placed at a scattering angle, $2\theta$, measures,

$$S_{\mathrm{Diff}}(Q_{\mathrm{el}}) = \int S(Q(E), E)dE, \tag{1}$$

where $S(Q, E)$ is the dynamic structure factor accessible by inelastic scattering, $E$ is the energy exchange in scattering, and $Q_{\mathrm{el}} = Q(E = 0)$. Because $Q$ varies with $E$ at constant $2\theta$, $S_{\mathrm{Diff}}(Q_{\mathrm{el}})$ is slightly different from $S(Q)$, for which the integration in Eq. (1) is made at a constant $Q$, making the PDF determined by diffraction measurement problematic, as discussed in Supplemental Material (SM) [20–22]. Furthermore, no attempt has been made to obtain information on dynamic electron correlation directly from the measurement of $S(Q, E)$.

In this Letter, we report on the observation of real-space dynamic electron correlation, using a real-space correlation function—specifically, the energy-resolved dynamic PDF, $g(r, E)$—which describes spatial electron correlation as a function of excitation energy [23]. This method has previously been employed to observe dynamic atomic correlations [24], e.g., the atomistic origins of relaxor ferroelectricity [25] and atom–atom correlations in superfluid $^4$He [26]. The $g(r, E)$ is obtained by Fourier transforming the $S(Q, E)$ over $Q$, by

$$g(r, E) = \frac{1}{2\pi^2 \rho} \int S(Q, E) \frac{\sin Qr}{Qr} Q^2 dQ \tag{2}$$

with $S(Q, E)$ measured over a wide $Q$-range to ensure accurate Fourier-transformation.

This conversion into the energy-resolved dynamic PDF facilitates the extraction of information on real-space dynamic correlations, whereas in the reciprocal space counterpart, $S(Q, E)$, such information is widely spread over $Q$ and buried under the signals for electron excitations. $S(Q, E)$ of electrons in light metals such as lithium, aluminum, and beryllium has been measured by using inelastic X-ray scattering (IXS) [17,19,27–31]. For a system of non-relativistic electron gas interacting with an electromagnetic field (X-rays), the Fermi's golden rule shows that the non-resonant differential scattering cross section is, within the Born approximation, proportional to $S(Q, E)$, given by

$$S(\boldsymbol{Q}, E = \hbar\omega) = \sum_{gf} \sum_{jj'} p_g \langle g| \mathrm{e}^{-\mathrm{i}\boldsymbol{Q}\cdot\boldsymbol{r}_j} |f\rangle \langle f| \mathrm{e}^{\mathrm{i}\boldsymbol{Q}\cdot\boldsymbol{r}_{j'}} |g\rangle \times \delta(E_g - E_f + \hbar\omega) \tag{3}$$

where $|g\rangle$ and $|f\rangle$ are the initial and the final states of the electron system, $p_g$ is the probability of state $|g\rangle$, and $E_g$ and $E_f$ are the energies of states $|g\rangle$ and $|f\rangle$, respectively [32,33]. The summation in terms of $j$ and $j'$ is over all electrons. By describing the response of the electron systems to an external electromagnetic field within a framework of linear response theory, the above equation is related to the response function of the system [34]:

$$S(\boldsymbol{Q}, \omega) = \frac{1}{\pi} \frac{1}{1 - \mathrm{e}^{-\beta\hbar\omega}} \chi''_{\rho^\dagger(\boldsymbol{Q})\rho(\boldsymbol{Q})}(\omega). \tag{4}$$

Here, $\chi''_{\rho^\dagger(\boldsymbol{Q})\rho(\boldsymbol{Q})}(\omega)$ is the imaginary part of the electron density response function, and $\rho(\boldsymbol{Q}, t)$ is the electron density fluctuation operator, which is the spatial Fourier transform of the electron density operator defined by $\rho(\boldsymbol{r}, t) = \sum_i \delta(\boldsymbol{r} - \boldsymbol{r}_i(t))$. Most of the previous IXS studies of electrons in solids use this description for the purpose of measuring single-particle excitations and collective excitations such as plasmons. However, as shown in SM, in this interpretation a random-phase approximation (RPA) is invoked to decouple the initial and final states, thereby the information on electron correlation is lost.

On the other hand, $S(\boldsymbol{Q}, \omega)$ can also be described in terms of the correlations in the density fluctuations in the system [35]. By using the Heisenberg time-dependent operator and by introducing the time-dependent correlation function of operators by $\sum_i p_i \langle i|A(0)B(t)|i\rangle \equiv \langle A(0)B(t)\rangle$, the dynamic structure factor is given by

$$\begin{aligned} S(\boldsymbol{Q}, \omega) &= \frac{1}{2\pi\hbar} \int_{-\infty}^{\infty} \mathrm{d}t\, \mathrm{e}^{-\mathrm{i}\omega t} \langle \rho(\boldsymbol{Q}, 0) \rho^\dagger(\boldsymbol{Q}, t) \rangle \\ &= \frac{1}{2\pi\hbar} \int G(\boldsymbol{r}, t) \mathrm{e}^{\mathrm{i}[\boldsymbol{Q}\cdot\boldsymbol{r} - \omega t]} \mathrm{d}\boldsymbol{r} \mathrm{d}t, \end{aligned} \tag{5}$$

where $G(\boldsymbol{r}, t)$ is the Van Hove correlation function [35],

$$G(\boldsymbol{r}, t) = \int \langle \rho(\boldsymbol{r} + \boldsymbol{r}', t) \rho(\boldsymbol{r}', 0) \rangle d\boldsymbol{r}'. \tag{6}$$

This representation provides a picture that the double-Fourier ($\boldsymbol{Q} \to \boldsymbol{r}$, $E \to t$) transformation of $S(Q, E)$ describes the dynamic density-density correlation of electrons. Finally, the energy-resolved dynamic PDF, $g(\boldsymbol{r}, E)$, is obtained by performing the Fourier transformation of $G(\boldsymbol{r}, t)$ over $t$:

$$g(\boldsymbol{r}, E) = \frac{1}{2\pi} \int G(\boldsymbol{r}, t) e^{-i(\frac{E}{\hbar})t} dt \tag{7}$$

The functions described in Eq. (4) and Eq. (5)–(7) encode equivalent information; however, projecting them onto different axes highlights different aspects of the function. Compared to the existing IXS works based on Eq. (4) that is suitable for describing the *response* of the system [36], this representation in real-space provides a more direct measure of the electron *correlation*. The information on electron correlation is included even in Eq. (4), but it is dispersed in $Q$ space and is hardly visible. It becomes recognizable only when it is brought to real space. The Van Hove correlation function is useful when excitations decay with time [37], but when excitations are long-living, the energy-resolved dynamic PDF is more informative.

To obtain the energy-resolved dynamic PDF for electrons, we use beryllium as a testbed because it is a simple metal with few core-electrons, a high electronic density [38] and a weak ionic potential that keeps electrons nearly free [39]. As this Letter aims to directly observe the dynamic correlation of electrons in beryllium in real-space, we measure $S(Q, E)$ of a polycrystalline beryllium plate at ambient conditions using medium-energy-resolution non-resonant IXS over wide ranges of $Q$ and $E$. The real-space dynamics are then extracted through the Fourier transformation of $S(Q, E)$ over $Q$. While IXS measurements with conventional X-ray sources [40–42] and synchrotron sources [18,27–31] have been reported for $S(Q, E)$ in various solids for a range of purposes, including studies of the linear response function, plasmon dispersion and its linewidth [19,27,28,41–43], and the effect of band structure [29] within a lower $Q$-range (well below 5 Å$^{-1}$), the real-space dynamic correlations of electrons have yet to be directly investigated.

One of the experimental challenges in the IXS of beryllium is its low X-ray absorption [44]. In transmission geometry, the low absorption results in an undesirably high background scattering originating from both the upstream and downstream of the sample. In reflection geometry, it leads to an exceptionally long footprint on the sample, complicating the normalization of the data. To mitigate systematic errors and artifacts specific to each experimental geometry, we carried out the IXS measurement in both transmission geometry and reflection geometry. The IXS measurement in transmission geometry was carried out using the spectrometer at 27ID-B in Advanced Photon Source (APS) at Argonne National Laboratory, and in reflection geometry using the spectrometer at P01 [45] in PETRA-III at DESY (See SM [20] for details of experimental setting and data corrections). The incident X-ray energy was 11.32 and 9.7 keV in the transmission and reflection geometry, respectively. A polycrystalline beryllium plate of thickness 1.0 mm was used as the sample. Scattering data were recorded within a horizontal scattering geometry for the transmission measurements and a vertical scattering geometry for the reflection measurement. The scattering angles ($2\theta$) spanned ranges of $8° < 2\theta < 90°$ (transmission) and $5° < 2\theta < 160°$ (reflection). At the APS, to maximize the scattering intensity at high angles in the horizontal setup, vertically polarized X-rays were produced by inserting a 250-μm-thick diamond 220 phase retarder [46]. The spectra were measured over $-4 < E/\mathrm{eV} < 120$ with an overall resolution of 0.7 eV at the APS and over $-10 < E/\mathrm{eV} < 1000$ with an overall resolution of 1.4 eV at PETRA-III. The measured spectra, $I(2\theta, E)$, were corrected for sample absorption and X-ray polarization [47] after the subtraction of elastic scattering, and were converted to $S(Q, E)$. The corrected spectra were then normalized using the $f$-sum rule to obtain the $S(Q, E)$ for a single valence electron [19].

*Results*—Usually, IXS measurements are conducted to determine the dispersion of excitations, focusing on the peak positions and their width. For this Letter, however, we sought to determine $S(Q, E)$ on an absolute scale over a wide range of $Q$, which demands a higher level of accuracy than usual. By comparing the data collected in both geometries, we confirmed that most of the observed spectral features are intrinsic to the material. In this Letter, we analyzed the energy spectra up to $E = 110$ eV, at which the X-ray Raman scattering is observed, resulting in $S(Q, E)$ for beryllium as shown in Fig. 1. A sharp plasmon peak is observed around $E = 21$ eV with little dispersion. Above $Q = 2$ Å$^{-1}$, the spectrum is dominated by Compton-like single-particle excitations. As numerous experiments have confirmed, plasmon excitations persist well beyond the critical wave vector $Q_c = 1.24$ Å$^{-1}$ [40]. At $Q < Q_c$, plasmons are damped mainly due to inter-band transitions [19]. Beyond $Q_c$ [48], damping intensifies as $Q$ increases, primarily due to Landau damping [49]. At much higher $Q$ ($Q > 2Q_F$, where $Q_F = 1.94$ Å$^{-1}$ is the Fermi wave vector for Be), plasmon becomes completely damped. In addition to the valence contribution, the $K$-edge structure, referred to as X-ray Raman scattering or Core-electron excitation, appears around 110 eV, with its intensity slightly increasing with $Q$ [43,50].

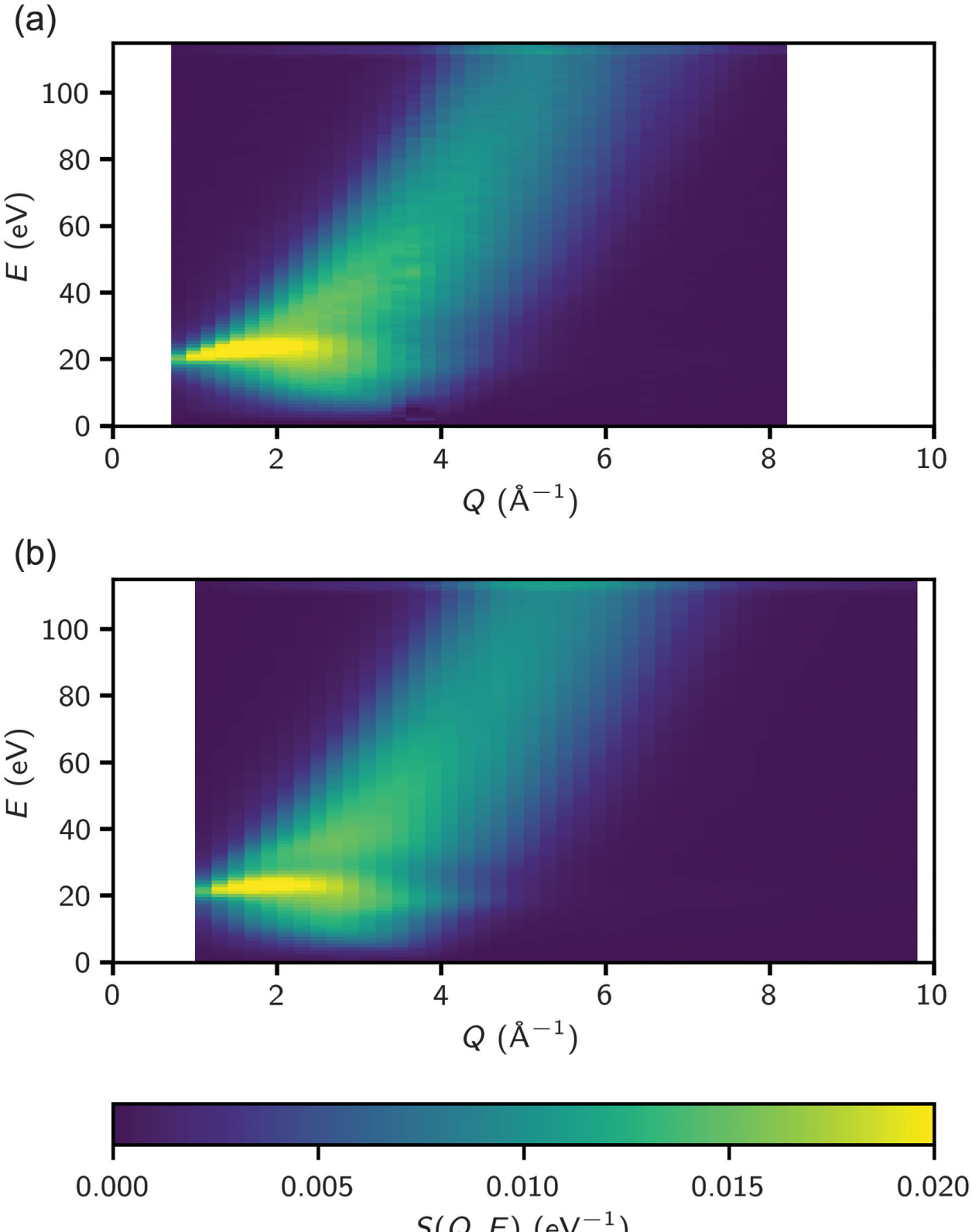

FIG. 1. Dynamic structure function, $S(Q, E)$, of electrons for beryllium measured at ambient condition. The $S(Q, E)$ was obtained from IXS measurements in (a) a transmission geometry at 27ID-B in APS and (b) a reflection geometry at P01 in PETRA-III. Their color scales are shown at the bottom.

The energy-resolved dynamic pair-distribution function, $g(r, E)$, which describes the spatial correlations of electrons at a specific excitation energy, can be obtained by the Fourier transformation of $S(Q, E)$ over $Q$ by Eq. (2). Figure 2 presents a two-dimensional plot of $g(r, E)$ for the electron system in polycrystalline beryllium. $g(r, E) = 0$ corresponds to the average electron density. At $r < 1$ Å, $g(r, E)$ is dominated by information related to the self-motion of electrons. In this Letter, we focus on a higher $r$-range ($r > 1$ Å) to examine the correlated dynamics of electrons. At $E < 40$ eV, which is relevant to the DFT calculations, there is a strong depression in electron density below 2 Å. This depression signifies the exchange-correlation hole. Its size, approximately 2 Å, is consistent with the size of the exchange-correlation hole for electrons in beryllium as we discuss below.

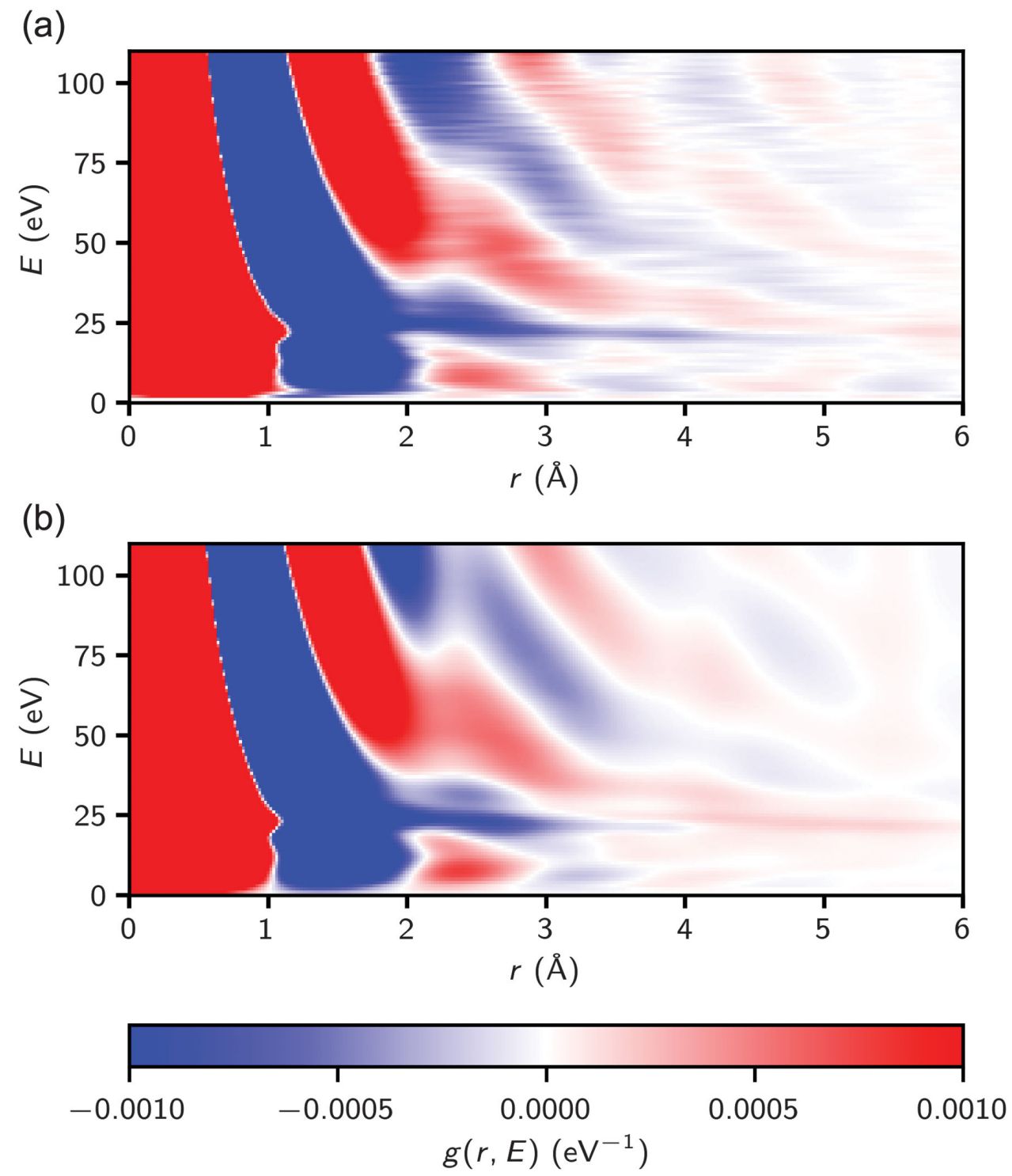

FIG. 2. Energy-resolved dynamic pair-distribution function, $g(r,E)$, of electrons in beryllium. The $g(r,E)$, determined by the IXS in (a) the transmission geometry and (b) the reflection geometry, is shown.

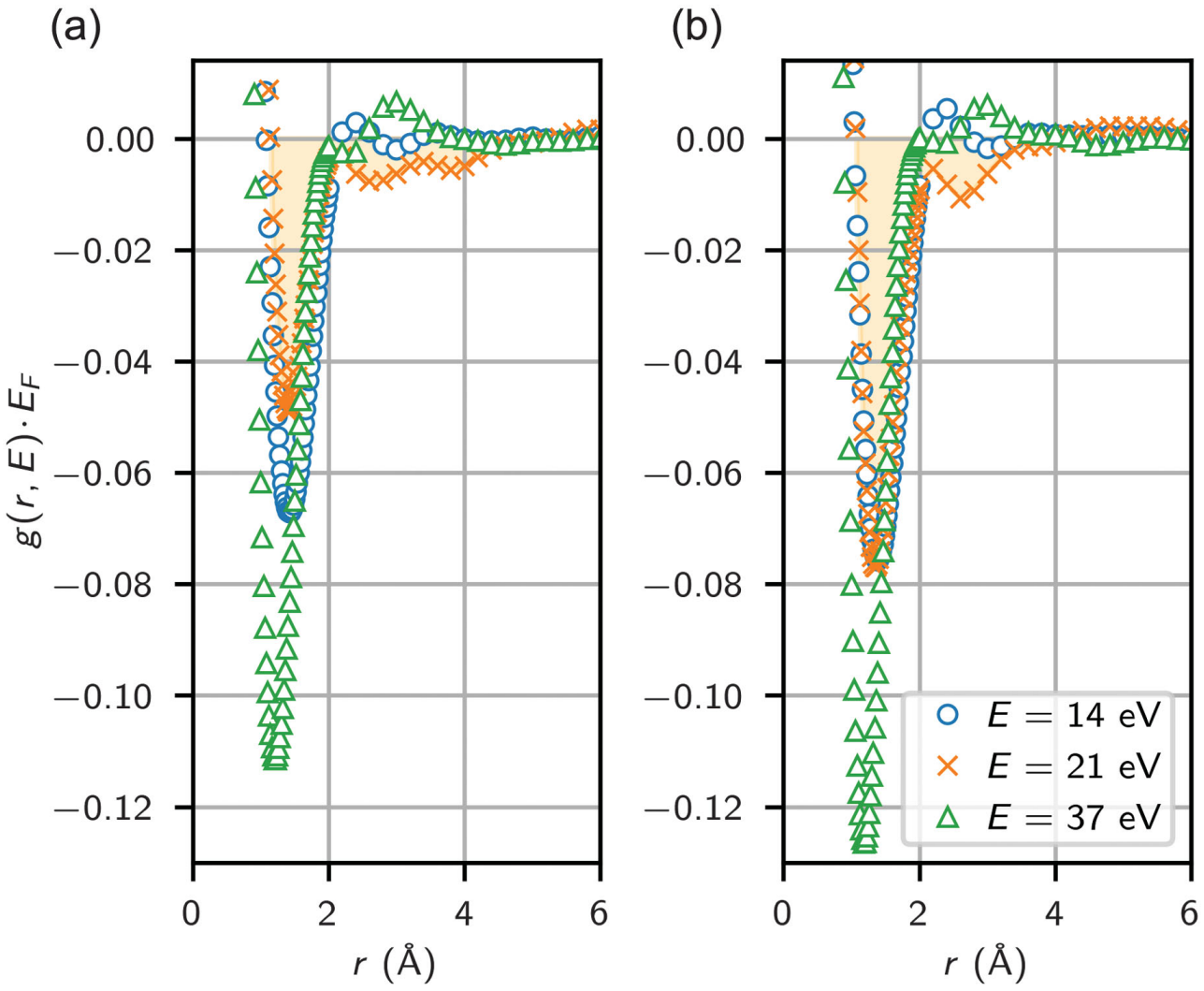

FIG. 3. Energy-sliced $g(r,E)$ of electrons in beryllium determined in (a) the transmission geometry and (b) the reflection geometry. The size of the exchange correlation hole for plasmon is different from that of average electrons; the energy-slices at $E = 14$ and 37 eV show the exchange-correlation hole for average electrons, which is roughly 2 Å, while at $E = 21$ eV, corresponding to plasmons, reveals a larger spatial extent of approximately 5 Å in (a) the transmission geometry and 4.2 Å in (b) the reflection geometry.

*Discussion*—This depression in $g(r,E)$ shows a unique behavior near the plasmon energies around 21 eV. Figure 3 shows $g(r,E) \cdot E_F$ above, near and below the plasma frequency. To facilitate comparison with other systems, the dynamic PDF is made dimensionless by multiplying it by $E_F$. At 14 and 37 eV, $g(r,E)$ oscillates around zero (average density) at distances beyond 2 Å, whereas at 21 eV, it is considerably depressed up to 4–5 Å. This result suggests that the exchange-correlation hole extends up to 4–5 Å in plasmon. Figure 4 shows $g(r,E)$ as a function of $E$ at different distances, highlighting how the density depression extends in energy. As the distance increases, the depression associated with the exchange-correlation hole becomes progressively shallower and narrower, eventually disappearing at around 5 Å. These findings suggest that the collective electron dynamics within plasmon affects electron correlation, leading to an extended exchange-correlation hole. Slight differences between the APS data and the PETRA-III data illustrate the extent of systematic errors arising from differences in experimental geometry and setup. However, these differences are insignificant and do not affect the overall conclusions.

There are intensity oscillations in $g(r,E)$ along $r$ as shown in Fig. 3, but these oscillations are intrinsic to the data and are not artifacts of termination errors. While all PDFs are inherently affected by termination effects due to the limited $Q$-range, these effects produce oscillations with a characteristic wavelength $\lambda_t = 2\pi/Q_{\max}$, where $Q_{\max}$ is the upper limit of the $Q$ range. In this Letter, $\lambda_t = 0.6 - 0.7$ Å, which is considerably shorter than the observed oscillation wavelength of ~1.4 Å, suggesting that these oscillations are not a consequence of termination errors. Instead, they are more likely attributable to the self-term, representing outgoing spherical waves of photoelectrons produced by Compton scattering. This subject will be discussed elsewhere.

The single particle excitations are suppressed below the plasmon energy. Thus, the exchange-correlation hole relevant to DFT is most appropriately evaluated in the energy range below the plasmon energy. As shown in Fig. 3, the exchange-correlation hole is well-defined with a size of approximately 2 Å and exhibits a strong depression of pair-density for distances less than 2 Å. However, the exchange-correlation hole appears to decrease in size with increasing energy (Fig. 2). This observation is misleading because $g(r,E)$ above the plasmon energy is dominated by the contribution from the self-term. The snapshot $g(r)$ is calculated by integrating $g(r,E)$ over $E$, and theoretically, the self-term at $r \neq 0$ should cancel out completely upon energy integration. Nevertheless, the strength of the self-term evident in Fig. 2 highlights that any experimental error could disrupt this cancellation, raising concerns about the reliability of the snapshot PDF obtained by diffraction measurements.

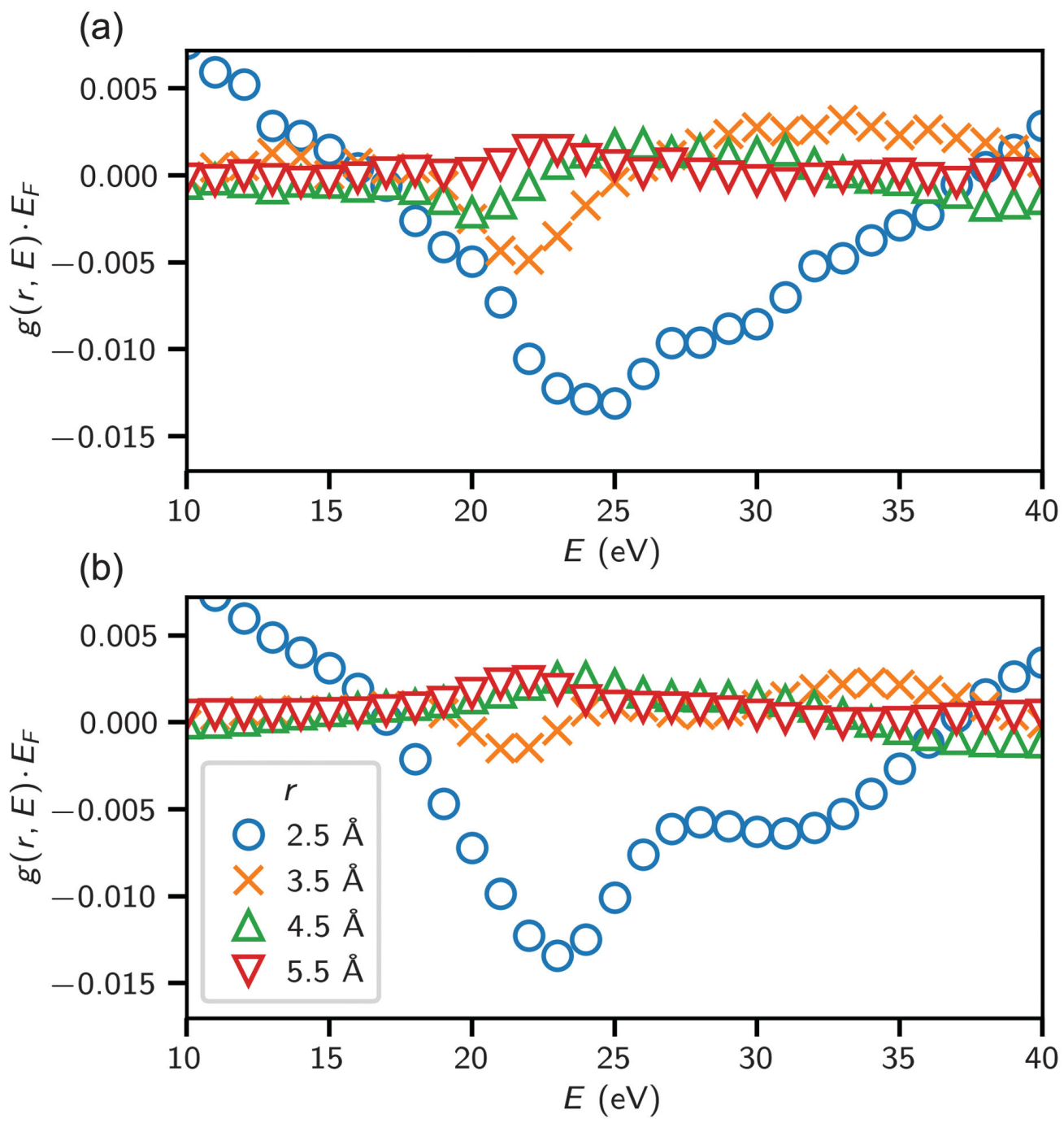

FIG. 4. Distance-sliced $g(r,E)$ of electrons in beryllium: This illustrates the energy dependence of the density depression in $g(r,E)$ at various fixed radial distances, $r$. As $r$ increases from 2.5 Å, the dip becomes progressively shallower and shifts to lower energy, eventually disappearing at 5.5 Å in (a) the transmission geometry and 4.5 Å in (b) the reflection geometry.

The size of the exchange-correlation hole, $r_{\mathrm{XC}}$, depends on the valence electron density, $\rho_{\mathrm{ele}}$. For beryllium ($\rho_{\mathrm{ele}} = 0.247$ Å$^{-3}$), $r_{\mathrm{XC}}$ is estimated to be approximately 2 Å, which closely matches our observed value (see SM for details). Besides, Perdew and Wang [51] developed an analytical method to construct the coupling-constant average pair-distribution function, $g(r,r')$, for a uniform electron gas with various density parameters. Their method was validated using Quantum Monte Carlo simulations [52,53] for both spin-polarized and spin-unpolarized systems. For beryllium, their result predicts $r_{\mathrm{xc}} \sim 2$ Å from the depletion in the exchange-correlation hole density given by $\rho_{\mathrm{XC}}(r,r') = \rho(r')[g(r,r') - 1]$, which is consistent with the results presented here (details in SM).

The extended exchange-correlation hole at the plasma frequency is likely attributed to plasmon dynamics. Within the plasmon, electrons move cooperatively with a reduced probability of impinging on each other, leading to a larger exchange-correlation hole. This observation may provide insight into an earlier observation on $^4$He. Through the energy-resolved dynamic PDF, it was found that the He-He distance in the Bose-Einstein (BE) condensate is longer by 10% compared to that in the non-condensate [26]. In the BE condensate, atoms are in the ground state without momentum, thus they do not impinge on each other just as in the plasmon, explaining the increased He-He distance. The cut-off momentum, $Q_c = 1.24$ Å$^{-1}$, characterizes the range of cooperative electron dynamics. Correspondingly, the associated wavelength, 5.1 Å, may define the spatial extent of the extended exchange-correlation hole at the plasma frequency.

*Conclusion*—Our Letter demonstrates that real-space dynamic correlations among electrons can be directly observed by IXS using the energy-resolved dynamic PDF. The results highlight distinct differences in correlation dynamics above, near, and below the plasma frequency. In particular, within the energy range relevant to the DFT calculations, the exchange-correlation hole is sharply defined with a size of ∼2 Å, consistent with theoretical estimations. This result contrasts with the conventional snapshot PDF approach obtained by diffraction without energy discrimination, which is prone to inaccuracies due to the possible errors at high energy transfers. Furthermore, the results highlight that the exchange-correlation hole associated with plasmons is more spatially extended, possibly due to cooperative electron dynamics intrinsic to the plasmon. This finding provides concrete evidence that electron correlation is affected by electron dynamics. The insights into real-space dynamic electron correlations, enabled by the energy-resolved dynamic PDF method, offer strong support for the DFT framework of electron correlation. These results also confirm that the measured dynamic structure function encodes information not only about single particle excitations but also about dynamic electron correlations. Although $S(\boldsymbol{Q},\omega)$, $G(\boldsymbol{r},E)$, and $G(\boldsymbol{r},t)$ contain equivalent information, they are projections onto different axes, revealing different features of the object. The underlying correlations are most readily visualized in real space, as demonstrated in this Letter.

This Letter initiates a field of direct experimental study of correlated electron dynamics, shedding critical new light on the broader field of condensed matter physics.

*Acknowledgments*—This work was supported by U.S. Department of Energy (DOE), Office of Science, Office of Basic Energy Sciences, Division of Materials Sciences and Engineering. This research used resources of the Advanced Photon Source, a U.S. DOE Office of Science User Facility operated for the DOE Office of Science by Argonne National Laboratory under Contract No. DE-AC02-06CH11357. We acknowledge DESY (Hamburg, Germany), a member of the Helmholtz Association HGF, for the provision of experimental facilities. Part of this research were carried out at PETRA III. Data was collected using P01 operated by DESY Photon Science. Beamtime was allocated for Proposal No. I-20230998. At the APS beamtime was allocated for Proposal No. 80535. ORNL is managed by UT-Battelle, LLC under Contract No. DE-AC05-00OR22725 with the U.S. Department of Energy.

*Data availability*—The data that support the findings of this article are openly available [54], embargo periods may apply.